\documentclass[twocolumn,preprintnumbers,amsmath,amssymb,superscriptaddress, altaffilletter, subeqn]{revtex4-1}

\usepackage{graphicx,wrapfig,lipsum}
\usepackage{times}
\usepackage{soul}
\usepackage[dvipsnames]{xcolor} 
\usepackage{mathtools}
\usepackage{siunitx}
\usepackage{comment}
\usepackage{mathrsfs}
\usepackage{physics}
\usepackage{feynmp} 
\usepackage{enumitem}

\def\ket#1{\mathinner{|{#1}\rangle}}

\newlength{\singlecolumn}
\renewcommand{\vec}[1]{\mathbf{#1}}

\makeatletter
\renewcommand*{\@fnsymbol}[1]{\ensuremath{\ifcase#1\or \dagger\or *\or  \ddagger\or
\mathsection\or \mathparagraph\or \|\or **\or \dagger\dagger
\or \ddagger\ddagger \else\@ctrerr\fi}}
\makeatother

\begin{document}

\title{Phonon-driven femtosecond dynamics of excitons in crystalline pentacene from first principles}

\author{Galit Cohen}
\affiliation{Department of Molecular Chemistry and Materials Science, Weizmann Institute of Science, Rehovot 7610001, Israel}
\author{Jonah B. Haber}
\affiliation{Department of Physics, University of California Berkeley, Berkeley, California 94720, USA}
\affiliation{Materials Sciences Division, Lawrence Berkeley National Laboratory, Berkeley, California 94720, USA}
\author{Jeffrey B. Neaton}
\affiliation{Department of Physics, University of California Berkeley, Berkeley, California 94720, USA}
\affiliation{Materials Sciences Division, Lawrence Berkeley National Laboratory, Berkeley, California 94720, USA}
\author{Diana Y. Qiu$^*$}
\affiliation{Department of 	
Mechanical Engineering and Materials Science, Yale University, New Haven, Connecticut 06520}
\author{Sivan Refaely-Abramson$^*$}
\affiliation{Department of Molecular Chemistry and Materials Science, Weizmann Institute of Science, Rehovot 7610001, Israel}

\begin{abstract}
Non-radiative exciton relaxation processes are critical for energy transduction efficiencies in optoelectronic materials, but how these processes are connected to the underlying crystal structure and its associated electron, exciton, and phonon band structures is poorly understood. 
Here, we present a first-principles approach to explore exciton relaxation pathways in pentacene, a paradigmatic molecular crystal and optoelectronic semiconductor. We compute the momentum- and band-resolved exciton-phonon interactions, and use them to analyse key scattering channels. We find that exciton intraband transitions on femtosecond timescales leading to dark-state occupation is a dominant nonradiative relaxation channel in pentacene. 
We further show how the nature of real-time propagation of the exciton wavepacket is connected with the longitudinal-transverse exciton splitting, stemming from crystal anisotropy, and concomitant anisotropic exciton and phonon dispersions. Our results provide a framework for understanding time-resolved exciton propagation and energy transfer in molecular crystals and beyond.

\end{abstract}
  
\maketitle

Molecular crystals are a class of materials broadly used as host systems for energy transfer processes, typically involving energetically excited states in which excitons---bound electron and hole pairs---serve as the main energy carriers~\cite{Ginley2011, Bradec2010}. Their light harvesting efficiency is regulated by exciton mobility and the dynamical processes it stems from. Well-explored examples are the acene-based crystals~\cite{Kronik2016}, and in particular pentacene, for which spectroscopic and microscopic measurements reveal unique non-radiative decay channels~\cite{Wilson.2013, Schnedermann.2019}. These decay processes are associated with optically-forbidden states due to spin, parity, and momentum. Long exciton lifetimes up to the order of nanoseconds, mainly associated with non-radiative relaxation processes, have been reported~\cite{Yost.2014, Rao2017, Casanova2018, Zhu2018, Sharifzadeh.2018}. The change in the spatiotemporal shape of the propagating light excitation, as seen in transient microscopy experiments~\cite{Zhu2017,Ginsberg2020}, allows for direct observation of the exciton time evolution patterns~\cite{Wan.2015, Schnedermann.2019}. These permit identification of the mechanisms dominating exciton relaxation processes leading to the elongated migration lifetimes, as well as the realization of underlying crystal structure and the associated optical selection rules. 

From a theoretical point of view, understanding non-radiative exciton relaxation mechanisms in molecular crystals requires a structure-sensitive analysis of the interactions dominating these processes, and in particular, their relation to crystal packing and symmetry. Given that the envelope function of optical excitations in these systems can span over several nanometers~\cite{Schuster.2007, Sharifzadeh.2013, Monahan.2015}, the role of crystal effects is expected  to be significant.  Indeed, crystal structure determines both the excitonic nature and its coupling to phonons
~\cite{Berkelbach.2014, Arago2015dynamics, Chang.2022, Seiler.2021, Alvertis.2020, berkelbach2017}. 
Numerous first principles calculations of excitons in solid pentacene have been  performed~\cite{sharifzadeh2012quasiparticle, Cudazzo2013, Coto.2015, Sharifzadeh.2013, Refaely2015, Rangel2016, cocchi2018polarized} using the \textit{ab initio} GW and Bethe–Salpeter equation (GW-BSE) approach~\cite{Hybertsen1987,Rohlfing2000}. Additionally, recent developments~\cite{Qiu2015, cudazzo2016exciton} allow for GW-BSE calculations of the exciton band structure~\cite{Refaely2017, Cudazzo.2015}, supplying further information on the exciton dispersion. Recently, several perturbative schemes to evaluate exciton-phonon interactions in a predictive manner, based on GW-BSE computations of excitons and density functional perturbation theory (DFPT)~\cite{Giustino.2017} or finite-displacement methods for phonons~\cite{Zacharias2016, Huang.2021}, were presented and applied to study phonon-assisted excitation, exciton relaxation and emission processes in layered and two-dimensional semiconductors~\cite{Chen2020, Paleari.2019, cannuccia2019theory,Antonius.2022, Zhang.2022, Chan2022}. 
While prior work has recently focused on how phonons can spatially localize excitons and renormalize excitation energies in organic crystals~\cite{Alvertis.2023}, studies of exciton scattering and dynamics in these systems are lacking.

In solid pentacene, recent calculations predicted two nearly-degenerate, low-lying singlet states, one optically bright (even parity state) and the second optically dark (odd parity state)~\cite{Refaely2017, cocchi2018polarized}. Such well-defined optical selection rules exist due to the presence of inversion symmetry, and the relative energies of these two exciton bands are dictated by the molecular monomers and their packing arrangement. Perturbative approaches based on GW-BSE suggest that this symmetry dictates coherent transitions between the low-lying singlet states and bi-triplets~\cite{Refaely2017, Altman.2022}, populating triplet states which largely increase the device efficiency~\cite{Folie.2018, Rao2017}. Furthermore, pentacene crystal anisotropy gives rise to a longitudinal-transverse (LT) splitting of the bright exciton branches, which was recently connected to ultrafast and highly anisotropic exciton transport in the short-time quasi-ballistic regime, prior to thermalization~\cite{Qiu2021}. It is important to understand the effect of the exciton LT splitting (and the crystal packing and anisotropy it stems from), coupled to phonon scattering, on exciton transport.   

\begin{figure}[h]
    \includegraphics[width=\linewidth]{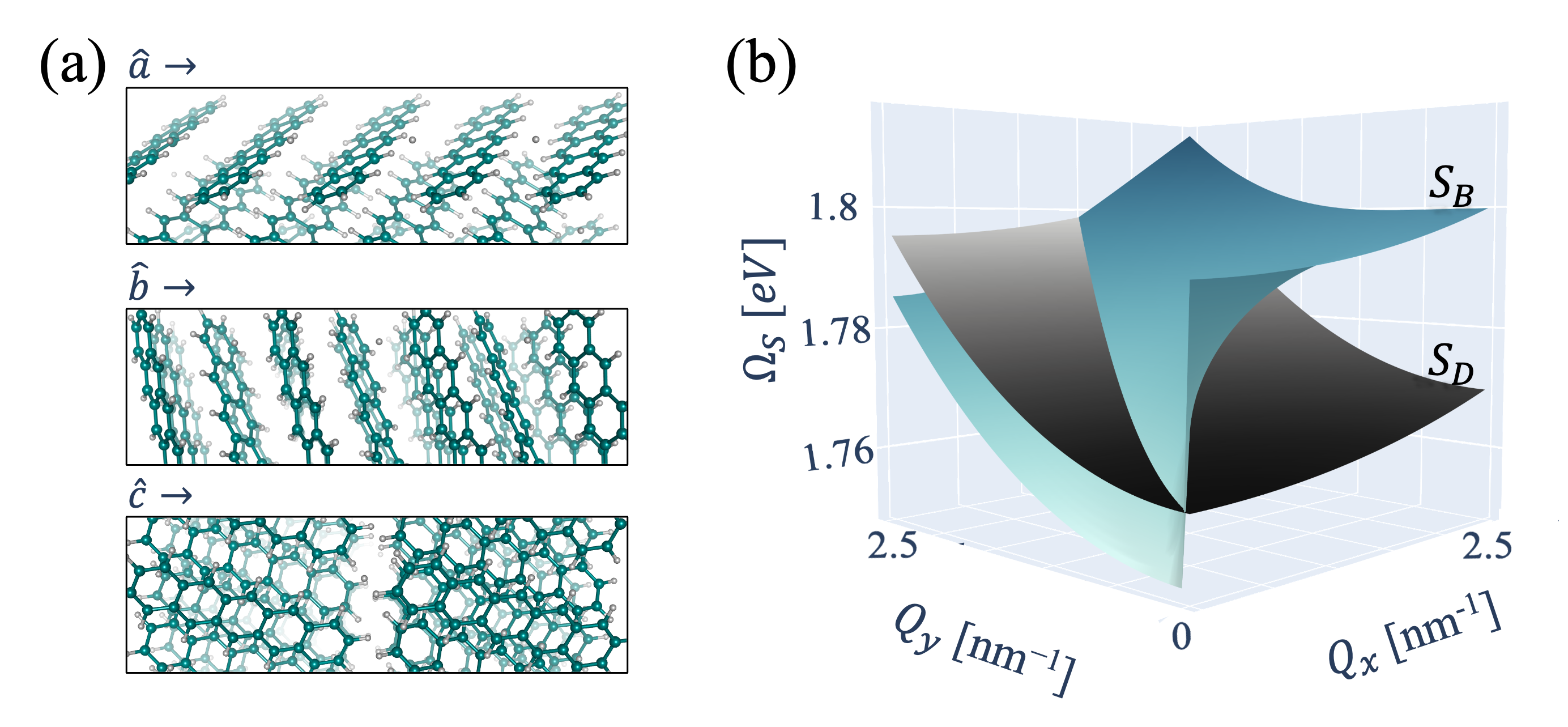}
    \vspace*{-10pt}
    \caption{(a) Molecular packing of the bulk pentacene crystal along different lattice vectors (carbon and hydrogen in teal and grey, respectively). (b) Exciton band structure of the low-lying bright ($S_B$, blue) and dark ($S_D$, black) singlet excitons. GW-BSE excitation energies are shown as a function of exciton momentum $\mathbf{Q}$ along the reciprocal $\mathbf{\hat{a}},\mathbf{\hat{b}}$ plane.} 
    \label{fig:basic_BSs}
\end{figure}

In this work, we present a first-principles approach, and apply it to pentacene, to study  exciton intraband and interband transitions and their role in the early stages of exciton transport and ultrafast occupation of long-lived states. Using a combination of \textit{ab initio} density functional theory (DFT), GW-BSE, and DFPT calculations, we identify key exciton relaxation channels and predict the ultrafast time evolution of a photoexcited exciton, and in particular the interplay between intraband scattering and interband decay into a nearly-degenerate dark state. Interestingly, we find these two processes have comparable timescales, a consequence of the large anisotropic exciton LT splitting. Moreover, the evolution of an exciton wavepacket delocalizes rapidly and anisotropically in a band-like transport regime, a further consequence of the exciton-phonon coupling and LT splitting. Our results suggest that exciton transport in pentacene involves phonon scattering already at very early stages, a property that is fully tunable through crystal structure. Our findings provide a basis for understanding the rapid exciton wavepacket spreading proceeding long-lived state occupation, as was recently observed in related experiments~\cite{Akselrod.2014,Wan.2015,Delor2020,Pandya.2021,Schnedermann.2019}.  

We study the crystalline pentacene in its bulk phase~\cite{Siegrist2007}, with a herringbone structure of van der Waals-stacked molecules, as shown in Fig.~\ref{fig:basic_BSs}(a) along the main three lattice directions (see SI for full details). 
Fig.~\ref{fig:basic_BSs}(b) shows the exciton band structure, computed with GW-BSE using the BerkeleyGW code~\cite{Deslippe2012berkeleygw, Qiu2015}. 
$\Omega_S$ is the exciton excitation energy, obtained from solving the BSE for electrons and holes with momentum difference $\mathbf{Q}_{x,y}$ at the $\mathbf{\hat{x}},\mathbf{\hat{y}}$ directions.
At $\vec{Q}=\Gamma$ the two low-lying singlet excitons are of opposite parities, one optically bright ($S_B$, blue) and the second optically dark ($S_D$, black). We color the exciton bands in Fig.~\ref{fig:basic_BSs}(b) accordingly. 
Importantly, the bright state exhibits a characteristic LT splitting at the $\Gamma$ point arising from long-range exchange interaction~\cite{Qiu2021}, previously explored analytically~\cite{Andreani.1988, Denisov1973, agranovich1966spatial}. This property introduces an angular dependency in the relative energy alignment between the two exciton bands, where the bright state has a higher (in the $\Gamma-X$ direction) or lower (in the $\Gamma-Y$ direction) excitation energy than the dark state. As a result, transitions between the bright and the dark states are expected, but their probability depends anisotropically on the exciton-phonon scattering. 

Phonon modes are computed using DFPT, with the unit cell lattice vectors fixed to the experimental values and internally relaxed with the PBE functional, a procedure shown to supply reliable phonon modes in organic crystals \cite{brown2016ab}. 
We combine the computed phonon and exciton states to evaluate the exciton-phonon matrix element via~\cite{Antonius.2022,Chen2020}:
\begin{equation}\label{eq:g_gab}
    \begin{split}
        \mathcal{G}_{SS'\nu} (\mathbf{Q},\mathbf{q}) &= \sum_{vcc'\mathbf{k}}[A_{vc\mathbf{k}}^{S\mathbf{Q}+\mathbf{q}}]^*g_{cc'\nu}(\mathbf{k}+\mathbf{Q},\mathbf{q})A_{vc'k}^{S'\mathbf{Q}}\\
                        &-\sum_{vv'c\mathbf{k}}[A_{vc\mathbf{k}}^{S\mathbf{Q}+\mathbf{q}}]^*g_{v'v\nu}(\mathbf{k},\mathbf{q})A_{v'c\mathbf{k}+\mathbf{q}}^{S'\mathbf{Q}}\;,
    \end{split}
\end{equation}
where $\mathcal{G}_{SS'\nu}(\mathbf{Q},\mathbf{q})$ encodes the coupling between two exciton states calculated within GW-BSE, $\ket{S(\mathbf{Q}+\mathbf{q})}$ and $\ket{S'(\mathbf{Q})}$, via a phonon with momentum $\mathbf{q}$ and mode $\nu$. $A_{vc\mathbf{k}}^{S\mathbf{Q}}$ are the amplitudes for $v,c$  the occupied (representing hole) and unoccupied (representing electron) states, respectively, and $\mathbf{k}$ is the electronic wavevector. $g_{c'c\nu}$ and $g_{v'v\nu}$ are the electron-phonon and hole-phonon coupling terms, calculated with DFPT. 
We define an exciton-phonon scattering time as~\cite{Antonius.2022,Chen2020}:
\begin{equation}\label{eq:tauS}
    \begin{split}
        \tau_{SS'}^{-1}=
        &\frac{2\pi}{\hbar} \frac{1}{N_q N_Q} 
        \sum_{\nu\mathbf{Q}\mathbf{q}}\abs{\mathcal{G}_{SS'\nu}(\mathbf{Q},\mathbf{q})}^2\rho(\Delta E_{SS'}(\mathbf{Q},\mathbf{q})-\hbar\omega_{\mathbf{q}\nu})\;,
    \end{split}
\end{equation}
where $N_q$ and $N_Q$ are the number of grid points in the phonon and exciton crystal momentum space, respectively, and the density of states is calculated via
\begin{equation}\label{rho}
    \begin{split}
        \rho(\Delta E_{SS'}&(\mathbf{Q},\mathbf{q})-\hbar\omega_{\mathbf{q}\nu})=\\
            &\big[(n_{\mathbf{q}\nu})\Tilde{\rho}(\Omega_{S(\mathbf{Q})}-\Omega_{S'(\mathbf{Q}+\mathbf{q})}+\hbar\omega_{\mathbf{q}\nu})\\
            &+(1+n_{\mathbf{q}\nu})\Tilde{\rho}(\Omega_{S(\mathbf{Q})}-\Omega_{S'(\mathbf{Q}+\mathbf{q})}-\hbar\omega_{\mathbf{q}\nu})\big]\;.
    \end{split}
\end{equation}
Here, $n_{\mathbf{q}\nu}$ is the phonon Bose-Einstein occupation function at temperature $T$, and we consider both phonon absorption and emission processes. $\Tilde{\rho}$ are Gaussian distributions with broadening $\sigma$. We denote the weighted scattering rate associated with each transition with mode and momenta resolution by $k_{SS'\nu}(\mathbf{Q},\mathbf{q})$, so that the scattering time introduced in Eq.~\eqref{eq:tauS} can be reexpressed as $\tau_{SS'}^{-1}= N_q^{-1} N_Q^{-1} \sum_{\nu\mathbf{Q}\mathbf{q}} k_{SS'\nu}(\mathbf{Q},\mathbf{q})$.

\begin{figure}[h]
\includegraphics[width=\linewidth]{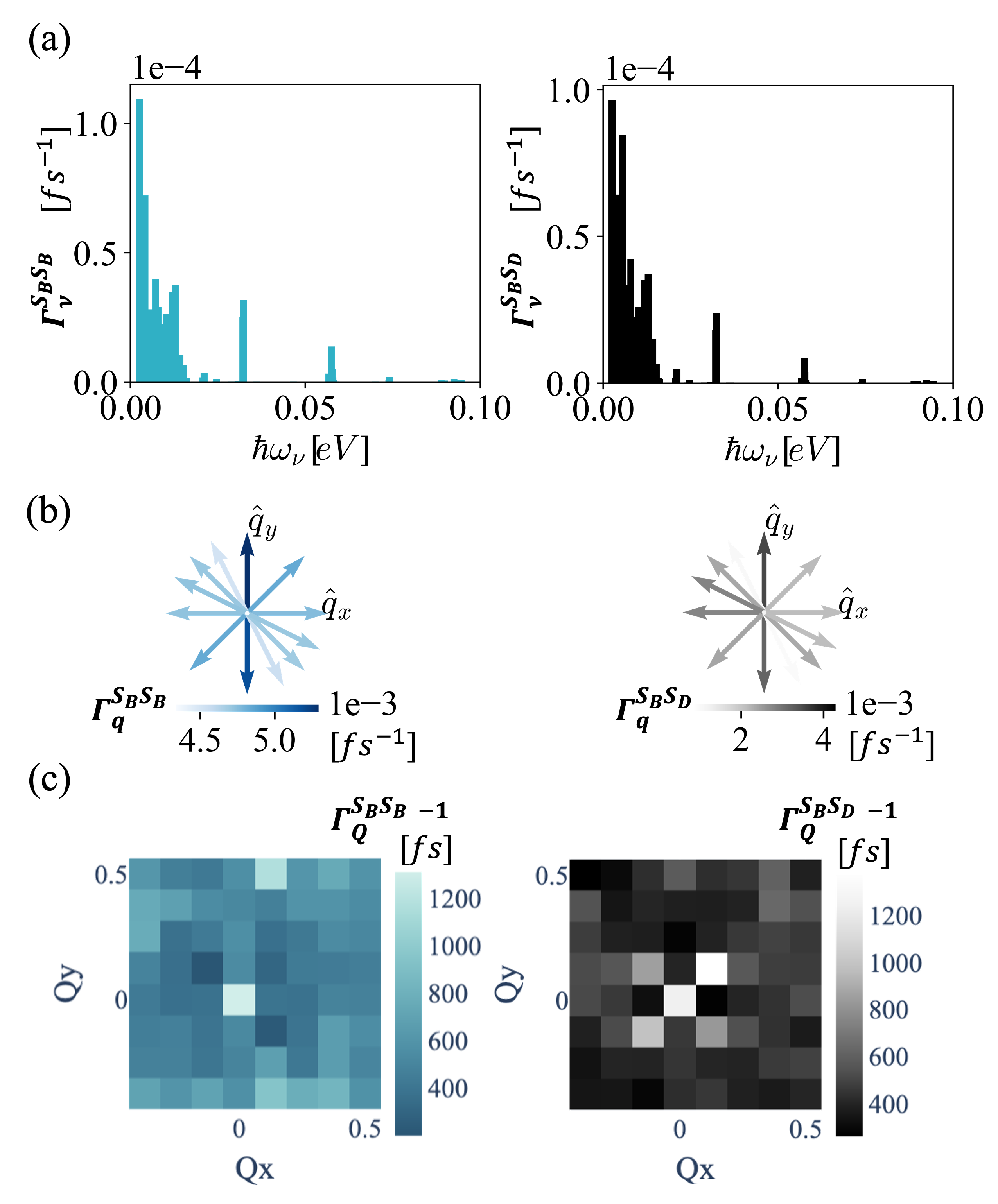}
    \vspace*{-10pt}
    \caption{Analysis of the computed exciton-phonon coupling for intraband (blue, left) and interband (black, right) transitions: (a) Computed exciton-phonon scattering rates as a function of phonon mode $\nu$. (b) Computed exciton-phonon scattering rates as a function of phonon momentum $\mathbf{q}$. (c) Computed exciton-phonon scattering times as a function of exciton momentum $\mathbf{Q}$.} \label{fig:gs}
\end{figure}

Fig.~\ref{fig:gs} shows our computed exciton-phonon scattering rates, for the two processes discussed above: intraband transitions from $S=S_B(\mathbf{Q})$ to $S'=S_B(\mathbf{Q}+\mathbf{q})$, namely within the optically-bright low-lying singlet state (blue, left); and interband transitions from $S=S_B(\mathbf{Q})$ to $S'=S_D(\mathbf{Q}+\mathbf{q})$, namely to the optically-dark state (black, right). 
In Fig.~\ref{fig:gs}(a) we further resolve the scattering rates over the phonon branch index
$\Gamma^{SS'}_\nu = N_q^{-1} N_Q^{-1} \sum_{\mathbf{Q}\mathbf{q}} k_{SS'\nu}(\mathbf{Q},\mathbf{q})$, and plot as a function of $\vec{q}$-averaged phonon energy, for $T=300~K$ and $\sigma=0.02~eV$ (this broadening parameter is compatible with the reciprocal space grid spacing and the energy difference between the exciton bands; see SI for further examination of the temperature and broadening effect). The bars along the phonon frequency axis demonstrate the dominance of low-frequency modes in both scattering channels. These include both acoustic and optical modes, and involve inter and intramolecular vibrations. 
The intermolecular modes are coupled to the more delocalized low-lying singlet excitons in this system, compared to localized excitons such as triplets which couple to intramolecular modes~\cite{Alvertis.2020,Seiler.2021,Hong-Guang.2020,Bakulin2016}. 
Fig.~\ref{fig:gs}(b) shows the scattering rates resolved per phonon momentum $\mathbf{q}$, $\Gamma^{SS'}_\mathbf{q} = N_Q^{-1} \sum_{\nu \mathbf{Q}} k_{SS'\nu}(\mathbf{Q},\mathbf{q})$. Arrows represent the phonon momentum direction and color stands for its associated rate, summed over all phonon modes and exciton momenta. 
Notably, we find larger contributions from phonons with momentum directed along $\Gamma-Y$, the transverse exciton direction, a property that plays an important role in the propagation dynamics as we show below. 

Fig.~\ref{fig:gs}(c) shows the inverse scattering rates resolved per exciton crystal momentum $\mathbf{Q}$, where $\Gamma^{SS'}_\mathbf{Q} = N_q^{-1} \sum_{\nu\mathbf{q}} k_{SS'\nu}(\mathbf{Q},\mathbf{q})$. Square colors indicate the scattering times from $S_B(\mathbf{Q})$ to states in the same band with different momentum (left) and to the dark band $S_D$ (right), summed over all momenta of the final exciton acquired due to the phonon momentum. 
The scattering times vary as a function of the initial exciton momentum, with weaker coupling around the $\Gamma$ point. In general, 
both intraband and interband transitions are determined by scattering processes from multiple momentum states. 
Importantly, we find comparable scattering times between exciton intraband transitions and the interband occupation of the dark state, with similar sensitivity to the scattered exciton momentum, associated with the characteristic exciton band structrue in this system. 
This unexpected result suggests that the early stages of exciton evolution involve an interplay between two main relaxation pathways: intraband transitions due to phonon scattering, and interband transitions to a dark state. Such dark states are long-lived, due to their robustness against radiative decay. Moreover, they strongly couple to bi-triplet excitons~\cite{Refaely2017,Neef.2023} which separate into triplet states, slowing down the exciton propagation ~\cite{Schnedermann.2019}. 

The rapid conversion of bright to dark states via phonon scattering 
is a clear consequence of the band crossing between these states, associated with the exciton LT splitting, together with the band alignment resulting from the molecular components and packing. It can thus be tuned via crystal structure design. For example, in the case of tetracene, the bright exciton band is higher in energy and does not cross the dark band. Such band alignment results in prolonged scattering times, with interband scattering times larger than intraband ones by an order of magnitude (see SI). 
We note that the computed scattering times should not be taken as the overall exciton lifetime, but rather as the characteristic scattering times of the individual decay pathways examined.
We further emphasize that the matrix elements composing the scattering times shown in Fig.~\ref{fig:gs} span a large range in magnitude, including times as short as $\sim30$~fs for strongly coupled states and up to few $\sim 100$~ps for weakly coupled ones. These are of the same order of magnitude
as reported coherent singlet-fission decay times of $\sim$70-100~fs, yet much shorter than observed non-coherent processes at the order of nanoseconds~\cite{Refaely2017,Wilson.2013, Rao2017, Seiler.2021, Neef.2023}. 

To connect our calculations with experimental observations, we study the effect of the computed intraband and interband scattering  on a propagating exciton wavepacket and explore its time-resolved evolution. The computed exciton-phonon coupling in this system is up to a few $meV$ (see SI), indicating a relatively weak interaction strength compared to the 
exciton bandwidth of $\sim~0.1$~$eV$. This suggests that the associated decay is due to weak interactions and supports the use of a perturbative treatment of the exciton scattering within a band-like transport regime. 
We  adopt a semi-classical kinetic equation form~\cite{Lifshitz1981} (see further derivation in the SI) in which  the time-resolved occupation of the optically-bright exciton, $n_{S_B}(\vec{Q},\vec{R})$, is evaluated in both real ($\mathbf{R}$) and reciprocal ($\mathbf{Q}$) space through 
\begin{equation} \label{eq:rate_eq_QR}
    \begin{split}
        \pdv{n_{S_B(\mathbf{Q},\mathbf{R})}}{t} = &-\pdv{n_{S_B(\mathbf{Q},\mathbf{R})}}{x}\cdot \pdv{\Omega_{S_B}}{\mathbf{Q}_x} -\pdv{n_{S_B(\mathbf{Q},\mathbf{R})}}{y}\cdot \pdv{\Omega_{S_B}}{\mathbf{Q}_y}\\
        &+ K_{scat}[n_{S_B(\mathbf{Q},\mathbf{R})}]\;.
    \end{split}
\end{equation}
The first two terms on the right-hand side of Eq.~\eqref{eq:rate_eq_QR} account for the ballistic wavepacket propagation, taking explicitly into account the computed GW-BSE exciton dispersion via the band velocities $\partial\Omega_{S_B}/\partial\mathbf{Q}$ and effectively coupling spatial and momentum coordinates along the time evolution.
The third term, $K_{scat}$, includes the computed exciton-phonon scattering terms presented above, and is expressed as: 
\begin{equation}\label{eq:rate_eq}
    \begin{split}
        K_{scat}[n_{S_B(\mathbf{Q},\mathbf{R})}]=
         \;&\;\;\; \sum_{\mathbf{q}} k_{S_B(\mathbf{Q}+\mathbf{q})S_B(\mathbf{Q})}\cdot n_{S_B(\mathbf{Q}+\mathbf{q})} \\
        & -\sum_{\mathbf{q}} k_{S_B(\mathbf{Q})S_B(\mathbf{Q}+\mathbf{q})}\cdot n_{S_B(\mathbf{Q})} \\
        & -k_{S_B(\mathbf{Q})S_D}\cdot n_{S_B(\mathbf{Q})}\\
        & -k^{rad}_{S_B(\mathbf{Q})}\cdot n_{S_B(\mathbf{Q})} \delta_{\mathbf{Q},\Gamma} \;.
    \end{split}
\end{equation}

Here, the first two scattering terms denote intraband transitions of the bright exciton with momentum $\mathbf{Q}$, in which the exciton $S_B(\mathbf{Q})$ is respectively created or destroyed. 
$k_{S_B(\mathbf{Q})S_B(\mathbf{Q}+\mathbf{q})}$ are the weighted scattering rates associated
with each momentum-resolved transition summed over all phonon modes, and can be expressed in terms of the scattering rates defined above $k_{S_B(\mathbf{Q})S_B(\mathbf{Q}+\mathbf{q})} \equiv \Gamma^{S_B S_B}_{\mathbf{q},\mathbf{Q}} = \sum_\nu k_{S_B S_B \nu}(\mathbf{Q},\mathbf{q})$.
The third term accounts for the rate of exciton decay due to interband scattering into the dark state, with the momentum of all final states summed through $S_D=\sum_{\mathbf{q}}S_D(\mathbf{Q}+\mathbf{q})$, thus $k_{S_B(\mathbf{Q})S_D}$ correspond to the above-defined $\Gamma^{S_B S_D}_\mathbf{Q}$. We exclude the reverse process, assuming that dark states go through rapid non-radiative decay (such as coherent coupling to bi-exciton pairs) before they can scatter back to a bright state. 
The last term represents the radiative decay from $S_B(\mathbf{Q}=\Gamma)$ to the ground state. We follow Refs.~\cite{spataru2005,Chen2019} and evaluate this term from the exciton oscillator strength and excitation energy (see SI for full derivation): 
\begin{equation}\label{eq:k_rad}
k^{rad}_{S_B}=\frac{2\pi e^{2}}{\hbar^{2}\mathrm{c}}\frac{\Omega_{S_B(\Gamma)}}{A_{uc}}\mu_{S}^{2}.
\end{equation}
Here $A_{uc}$ is the unit cell planar area, $\Omega_{S_B(\Gamma)}$ is the exciton energy at $\mathbf{Q}=\Gamma$, $\mathrm{c}$ is the speed of light and $\mu_s^2$ is the modulus square dipole strength of the exciton divided by the number of $\mathbf{k}$-points. This leads to a recombination lifetime of $\sim$3.5~ps, in agreement with a previously-reported trend of relatively low radiative rates and small measured radiative signal in pentacene crystals~\cite{Jundt.1995,Wilson.2013,Marciniak.2007}.\par

\begin{figure}[t]
    \includegraphics[width=\linewidth]{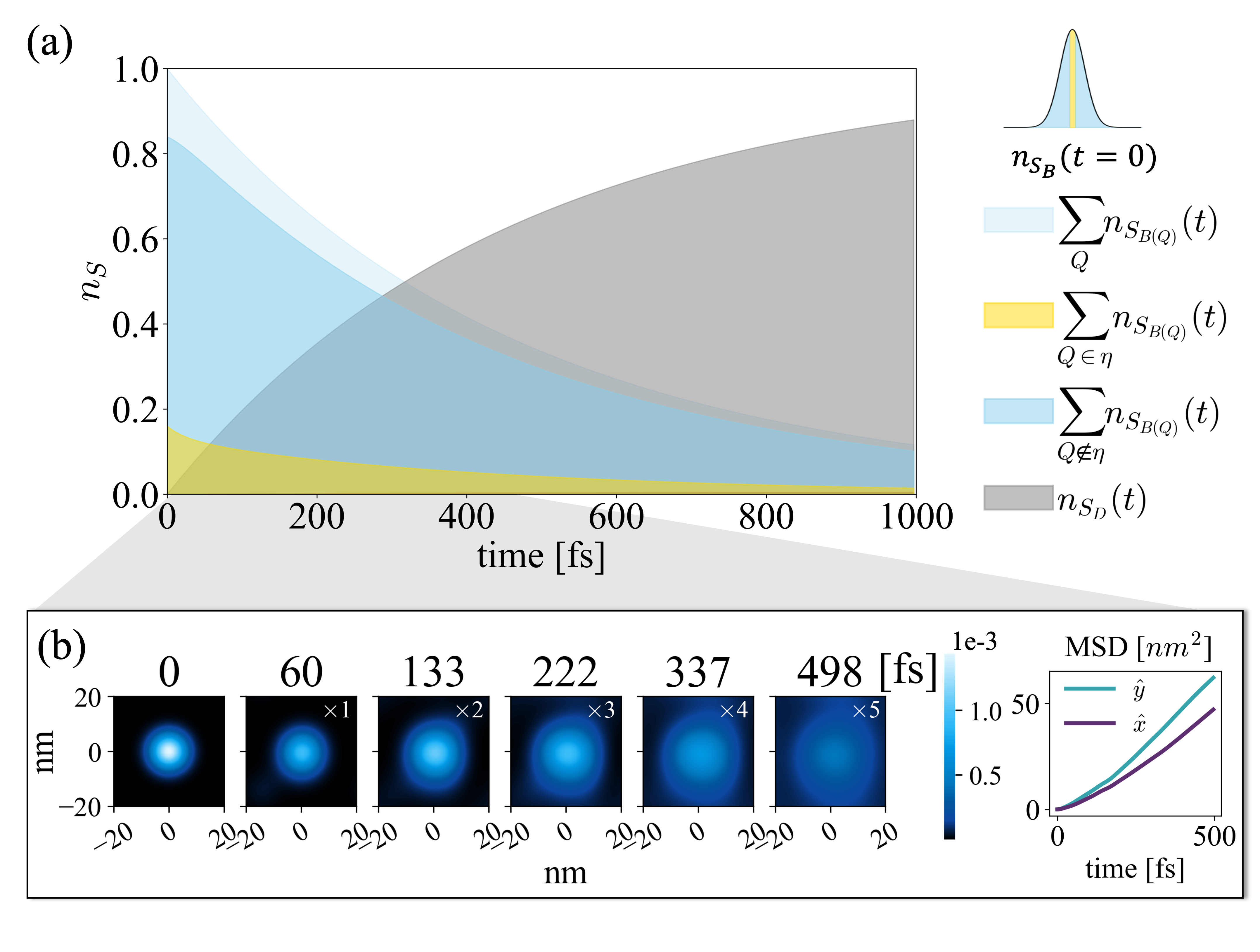}
    \vspace*{-10pt}
    \caption{(a) Exciton wavepacket occupation, computed from Eq.~\eqref{eq:rate_eq_QR} for the first 1~ps after excitation, with the scattering rates constructing Fig.~\ref{fig:gs}. The exciton evolution consists of the change in populations of $S_B$ state at $\mathbf{Q}\approx\Gamma$ (within the area $\eta$ around $\Gamma$, including the points where the initial occupation is higher than $0.8 n_{S_B}(t=0)$) (yellow), $S_B$ state at the remainder of $\mathbf{Q}$-space (blue), total $S_B$ state population (light blue), and $S_D$ state at all $\mathbf{Q}$-points (grey). (b) Spatial propagation of the wavepacket during the first 500~fs. Blue distribution shows the bright state occupation, normalized and amplified with respect to the initial step. The mean squared displacement along the $\mathbf{\hat{x}}$ (purple) and $\mathbf{\hat{y}}$ (teal) are shown on the right.}
    \label{fig:rate_eq}
\end{figure} 

Having computed all components of Eq.~\eqref{eq:rate_eq_QR} we calculate the real-time dynamics of the photoexcited exciton, within the approximations discussed above, starting from an initial Gaussian exciton wavepacket~\cite{Qiu2021} centered around the zero momentum and spatial positions, with broadening of 2~nm$^{-1}$ and 3~nm in $\mathbf{Q}$ and $\mathbf{R}$ space, respectively. 
These initial conditions are sufficient to observe the momentum-resolved processes that dominate its early-time evolution (see SI for further analysis). 
Fig.~\ref{fig:rate_eq}(a) shows our computed exciton population as a function of time after excitation. We differentiate the initial occupation of the $S_B(\mathbf{Q})$ state, shown schematically in the legend, with the populations of $\mathbf{Q}$-points near $\Gamma$ (yellow), the tails far from it (blue), and their corresponding sum (light blue). The area near $\Gamma$ (denoted by $\eta$) includes the points where the initial occupation is higher than $0.8 n_{S_B}(t=0)$. Due to the interband scattering, a growing occupation of the dark state follows (grey), competing with the non-radiative intraband scattering. Radiative recombination is negligible at these timescales and in the presented population range. 
At early times of the exciton evolution, up to 200~fs, the initial population mainly decays within the bright state via intraband transitions; after which the decay to the $S_D$ state becomes dominant. After 1~ps, the dark state occupies $\sim$80\% of the exciton population.

Fig.~\ref{fig:rate_eq}(b) shows the spatial propagation in real space, presented with normalized intensity amplified with respect to the initial time step (the transition to the dark state leads to a strong decrease in the non-normalized signal, see SI). The mean squared displacement (MSD) is also shown along the $\mathbf{\hat{x}}$ (purple) and $\mathbf{\hat{y}}$ (teal) directions. 
At $\sim$100-200~fs the Gaussian wavepacket starts to delocalize along both the $\mathbf{\hat{x}}$ and $\mathbf{\hat{y}}$ directions. This is a direct outcome of the exciton LT splitting, as we observed previously for the ballistic regime~\cite{Qiu2021} (further analysis is given in the SI). However, unlike in the ballistic regime, the exciton-phonon scattering leads to enhancement of the wavepacket broadening, as indicated by the MSD analysis, with a preferred directionality of the propagation along $\mathbf{\hat{y}}$ stemming from the nonuniform exciton-phonon scattering in addition to the exciton dispersion anisotropy demonstrated above. 

We note three important aspects revealed by our calculations in the propagation picture. First, the wavepacket broadening is anisotropic, directly manifesting the crystal anisotropy through both the LT exciton splitting and phonon scattering, connecting the propagation pattern to the crystal structure. Second, the increase in exciton spread happens at comparable time scales to dark state occupation, opening the door to tuning the various scattering times via interplay between the bright and dark bands alignment, controlling the inter and intraband scattering channels. Third, the computed change in the wavefunction broadening is non linear, due to the strong coupling between the LT splitting of the exciton band and the phonon scattering. Still, one can assume linearity for the region above 200~fs, where the ballistic effects reduce and the phonon scattering dominates. At this region we find diffusion coefficients of 0.42~$\frac{cm^2}{s}$ and 0.55~$\frac{cm^2}{s}$ along the $\mathbf{\hat{x}}$ and $\mathbf{\hat{y}}$ directions, respectively. 
At later times, the occupation of the dark state starts to dominate the exciton transport. 

To conclude, we presented a theoretical scheme to calculate time-resolved exciton evolution in crystals, applied to crystalline pentacene and offering a direct relation between exciton dynamics and crystal structure. Our approach is based on an \textit{ab initio} evaluation of the exciton-phonon scattering rates, allowing us predictive assessment of  scattering probabilities for intraband thermalization of the photoexcited state and its transition to a dark state, a consequence of the crossing between the dark exciton band and the bright LT split exciton band. This band-crossing results in unexpected exciton evolution, coupling ballistic effects due to the LT exciton splitting with phonon scattering and is manifested in rapid exciton wavepacket spreading before the occupation of long-lived dark state. This interplay is expected to dominate the efficiency of energy transfer through effective activation of non-radiative relaxation pathways, which is tunable through both crystal packing and molecular compositions. 
Our results allow for a detailed understanding of the relation between crystal packing, anisotropy, and symmetry to the early stages of exciton transport, setting the grounds for a predictive, first-principles description of the underlying dynamical processes dominating efficient energy transfer in organic semiconductors.

\textbf{Acknowledgments:} We thank Lev Melinkovsky, Mikhail Glazov, Felipe da Jornada, Alexey Chernikov, Andr\'{e}s Montoya-Castillo, and Aaron Kelly for insightful discussions. G.C. acknowledges a SAERI Doctoral Fellowship.
D. Y. Q. is supported by an Early Career Department of Energy Award Number DE-SC0021965. S. R. A. is an incumbent of the Leah Omenn Career Development Chair and acknowledges a Peter and Patricia Gruber Award and an Alon Fellowship.
The development of a propagating wavepacket model for excitons was supported by the U.S. Department of Energy, Office of Science, Office of Basic Energy Sciences grant DE-SC0021965. 
The project has received further funding from the European Research Council (ERC), Grant agreement No. 101041159, and an Israel Science Foundation Grant No. 1208/19.
Computational resources were provided by the National Energy  Research Scientific Computing Center (NERSC), as well as the ChemFarm local cluster at the Weizmann Institute of Science.

\end{document}